\documentclass[aps,prb,twocolumn,showpacs,showkeys,groupedaddress,amsmath,amssymb]{revtex4}

\usepackage{graphicx}
\usepackage{color}
\usepackage{bm}     
\usepackage{ulem}
\usepackage{amsmath}

\begin{document}

\title{Magnetoelastic coupling in $\gamma$-iron}

\author{S.~V. Okatov}
\affiliation{Institute of Quantum Material Science, 620107, Ekaterinburg, Russia}

\author{Yu.~N. Gornostyrev}
\affiliation{Institute of Metal Physics, Russian Academy of
Sciences, Ural Division, Ekaterinburg, 620219}
\affiliation{Institute of Quantum Material Science, 620107, Ekaterinburg, Russia}

\author{A.~I. Lichtenstein}
\affiliation{Institut f\"ur Theoretische Physik, Universit\"at
Hamburg, Jungiusstrasse 9, 20355, Hamburg, Germany}

\author{M.~I. Katsnelson}
\affiliation{Radboud University Nijmegen, Institute for Molecules and
Materials, 6525AJ, Nijmegen, the Netherlands}


\begin{abstract}
Exchange interactions in $\alpha$- and $\gamma$-Fe are
investigated within an {\it ab-initio} spin spiral approach. We
have performed total energy calculations for different magnetic
structures as a function of lattice distortions, related with
various cell volumes and the Bain tetragonal deformations. The
effective exchange parameters in $\gamma$-Fe are very sensitive to
the lattice distortions, leading to the ferromagnetic ground state
for the tetragonal deformation or increase of the volume cell. At
the same time, the magnetic-structure-independent part of the
total energy changes very slowly with the tetragonal deformations.
The computational results demonstrate a strong mutual dependence
of crystal and magnetic structures in Fe and explain the
observable ``anti-Invar'' behavior of thermal expansion
coefficient in $\gamma$-Fe.
\end{abstract}


\maketitle

\section{Introduction}

Iron-based alloys are still among the most important
industrial materials. The thermodynamic properties and
mechanism of phase transformations in these materials have been
discussed intensively last years \cite{Cahn-book}. Nevertheless, the
fundamental properties of iron have not been completely understood
up to now. Main difficulties are related with a non-trivial
combination of the itinerant and localized behavior
and correlation effects of
$3d$-electrons determining electronic, magnetic and
structural properties of iron
\cite{Herring1966,Vonsovsky1974,VonsKatsTref1993FMM,LichtKatsKotliar2001PRL,
KatsLicht2010PRL}. It is commonly accepted now that magnetic
degrees of freedom play a crucial role in the stability of different iron phases
\cite{9,10,11,danil} which makes the situation even more
complicated. Interplay between magnetic and lattice degrees of freedom in
different crystallographic phases of iron remains still unresolved problem.

One of the most complicated example related $\gamma$-phase of iron
with highly frustrated magnetic structure. There are many magnetic
configurations of $\gamma$-Fe with almost the same total energies
and the ground state is crucially depends on the value of
lattice parameters \cite{antropov-kats,abrikos,hafner}. The
sensitivity to dilatation has been studied in detail by many groups
\cite{Abrikosov,mryasov2,Ruban} in the context of a so called
moment-volume instability \cite{Wasserman}. At the same time, the
energy dependence on the tetragonal deformation which is
closely related with the Bain deformation path of
$\alpha-\gamma$ phase transformation also deserves a serious attention.
We discussed this issue in our previous work \cite{okatov} and
found that the transition of $\gamma$-Fe to the ferromagnetic
state can trigger the martensitic transformation without
noticeable energy barriers. In more detail, the effect of
tetragonal deformations on magnetism and vice versa was discussed
in relation with the Invar behavior observed in Fi--Ni alloys
\cite{Wasserman,Mohn-Khmelevski}. A magnetoelastic
spin-lattice coupling plays also an important role in structural
phase transitions in $\gamma$-Mn \cite{cumn} and Cr-based alloys
\cite{chromium}, as well as in the magnetic shape-memory alloy
Ni$_2$MnGa \cite{ni2mnga}. A soft-mode phonon behavior, as a
precursor of the $\gamma \rightarrow \alpha$ transformation, was
recently observed in Fe--Ni alloys \cite{tsunoda08}.

In contrast with Fe--Ni alloys, the equilibrium $\gamma$-phase in
pure Fe exists only at high temperatures $T>1200$~K where thermal
fluctuations are very strong and magnetic moments are disordered.
Observation of the so-called ``anti-Invar'' behavior of
$\gamma$-Fe \cite{Acet} can be related with the fact that the spin--lattice
coupling is strong enough to affects the thermodynamic properties
up to very high temperatures.

In this paper we investigate quantitatively a variation of the exchange
parameters in $\alpha$- and $\gamma$-Fe as functions of tetragonal
Bain-deformations and dilatation. Whereas the sensitivity of the
exchange parameters to the dilatation has been studied previously
\cite{Abrikosov,mryasov2,Ruban} an information about the tetragonal
deformations have been missing until now. Based on the calculated magnetic
exchange data we discuss the origin of the anti-Invar behavior of $\gamma$-Fe.

\section{Computational approach}

The standard approach to study magnetic properties of itinerant-electron
transition-metal systems related with the mapping of density functional total energies
on the effective classical Heisenberg model:
\begin{equation}
H_{ex} = - \sum_{i<j} J_{i,j} {\bf e}_i {\bf e}_j
\label{heisen}
\end{equation}
were ${\bf e}_i$ is the unit vector in direction of the magnetic moment
at site $i$ \cite{exchange,liecht}. In this notation the value of
on-site atomic magnetic moments $M_i$ is included into the exchange
parameters $J_{i,j}$. Therefore the total energy of the system is a sum
of a magnetic-structure independent contribution $E_0$ and the
``Heisenberg-exchange'' part: $E=E_0+H_{ex}$, where $E_0$ is a function
of deformations and the magnitude of local moments: $E_0(\Omega,c/a,M)$.
A similar decomposition was used earlier in Ref.~\onlinecite{kubler}.
One should stress that $E_0$ is dependent on the values of magnetic
moments $M_i$ and is therefore essentially different from the energy of
a non-spin-polarized state, which attribute is zero all magnetic
moments: $M_i=0$.

There are two main approaches to the mapping onto magnetic Hamiltonian.
An analytical scheme is based on the use of so-called ``magnetic force theorem''
\cite{exchange,liecht}, when the exchange interactions are
obtained from variations of the total energy with respect to
infinitesimal deviations of the magnetic moments from a
collinear state. In this paper we use more accurate numerical method based on the
density functional calculations of the spin spiral magnetic
structures where the neighboring magnetic moments are rotated
relative to each other by a finite angle (for review, see
Ref.~\onlinecite{Sandratskii_Adv_Phys}). This scheme includes
a spin- and charge-density relaxation for large moment fluctuations.
The energy per atom of the spin spiral with the wave vector ${\bf Q}$ can be presented
as:
\begin{eqnarray}
E({\bf Q}) && = E_0 - \dfrac{1}{N} \sum_{i<j} J_{i,j} \exp (i{\bf Q}\cdot {{\bf R}_{i,j}})
\nonumber \\
 && = E_0 - \sum_{n} Z_n J_{n} \exp (i{\bf Q}\cdot {{\bf R}_{n}}),
\label{h-ss}
\end{eqnarray}
where $N$ is the number of magnetic atoms, $Z_n$ is the number of the $n$-th
nearest neighbor atoms, $E_0$ is a magnetic-structure-independent
contribution to the total energy of the system, ${\bf R}_{i,j}$ is the
vector connecting sites $i$ and $j$, $n$ labels the coordination shell.
The exchange parameters $J_n$ can be found from Eq. (\ref{h-ss}) by
using the discrete Fourier transformation:
\begin{eqnarray}
J_n = -\dfrac{1}{K} \sum_{k} E({\bf Q}_k)
\exp(i{\bf Q}_k\cdot {{\bf R}_{n}}),
\label{Jn}
\end{eqnarray}
where the summation runs over a regular ${\bf Q}$-vector mesh in Brillouin zone with the
total number of points $K$. As follows from Eq.(\ref{Jn}) the value
$E_0$ is the average value of spin spiral energies over all ${\bf Q}_k$,
\begin{eqnarray}
E_0= \dfrac{1}{K} \sum_{k} E({\bf Q}_k).
\label{E0}
\end{eqnarray}
In principle, one can find the dependence of total energy on magnitude of $M$
within a constrained moment spin-spiral calculations, but this lay beyond
a scope of present paper.
A parameter of total exchange energy
\begin{eqnarray}
J_0=\sum_{n}Z_n J_n
\label{J0}
\end{eqnarray}
characterizes a ferromagnetic contribution to the total energy. Note
that the decomposition of the total energy used in
Ref.~\onlinecite{kubler} differs from that used in this work by
a shift by $J_0$.

In general, the exchange parameters found from the planar spin
spiral calculations and from the magnetic force theorem are
different and only the value of a spin stiffness constant should be
the same \cite{mik04}. Note that parameters obtained
by the use of infinitesimal spin deviations \cite{exchange,liecht} give a
correct description of a magnon spectra, while parameters found
from a direct calculation of the spin spiral total energies are
supposed to be more accurate for descriptions of thermodynamic properties
\cite{mik04}. The difference of $J_n$ obtained within these two
approaches characterizes a non-Heisenberg character of magnetic
interactions which is expected for itinerant magnets such as iron
\cite{tur_licht_kats90}. Another manifestation of
the non-Heisenberg behavior related with the fact that the magnitude of the magnetic
moments dependent on the spin spiral wave vector
$\mathbf{Q}$. Therefore, the values $J_n$ obtained in the
framework of spin spiral approach are considered as {\it
effective} exchange parameters.

The total energy calculations of Fe with spin spirals magnetic structure
is performed using VASP (Vienna Ab-initio Simulation Package)
\cite{kresse1,kresse2,kresse3} with first-principle pseudopotentials
constructed by the projected augmented wave method (PAW) \cite{blochl}.
Following an experience on non-collinear magnetic investigation\cite{hafner}
we employed the generalized gradient approximation (GGA) for the density
functional in a form by Perdew and Wang (1991)\cite{27} with the
spin-interpolation \cite{28}.  The PAW
potential without core states and with energy mesh cutoff 530 eV, and
the uniform $k$-point 12$\times$12$\times$12 mesh in the Monkhorst-Park
scheme \cite{MP} with 1728 $k$-points are used. The calculations are
done for a single-atom unit cell subjected by two types homogeneous
deformations, namely, dilatation (a change of the volume for a fixed $c/a$
ratio) and tetragonal ones (a change of $c/a$ ratio at a fixed volume).
For given lattice parameters, the energy set $E({\bf Q}_k)$ is
calculated on a uniform 16$\times$16$\times$16 mesh and the Fourier
transformation (Eq.~\ref{Jn}) is used to determine the exchange
parameters $J_n$.

\begin{figure}
\resizebox*{0.98\columnwidth}{!}{\rotatebox{0}{\includegraphics{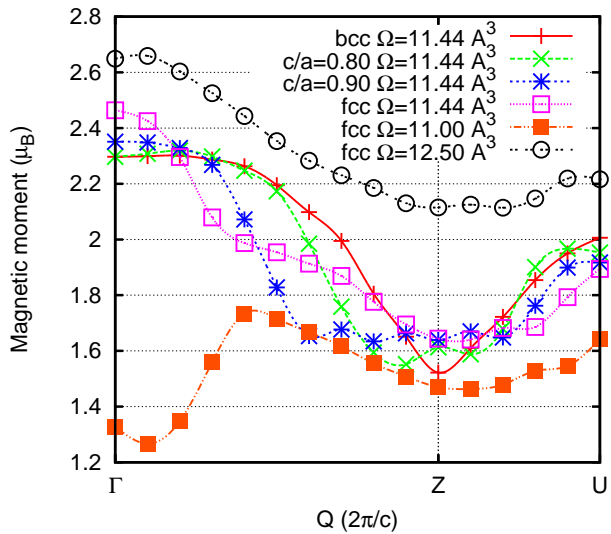}}}
\caption{(Color online) The magnetic moment of iron as function of the
spin spiral vector {\bf Q}  and the lattice deformations.}
\label{fig_spiral_m}
\end{figure}

\begin{figure}
\resizebox*{0.98\columnwidth}{!}{\rotatebox{0}{\includegraphics{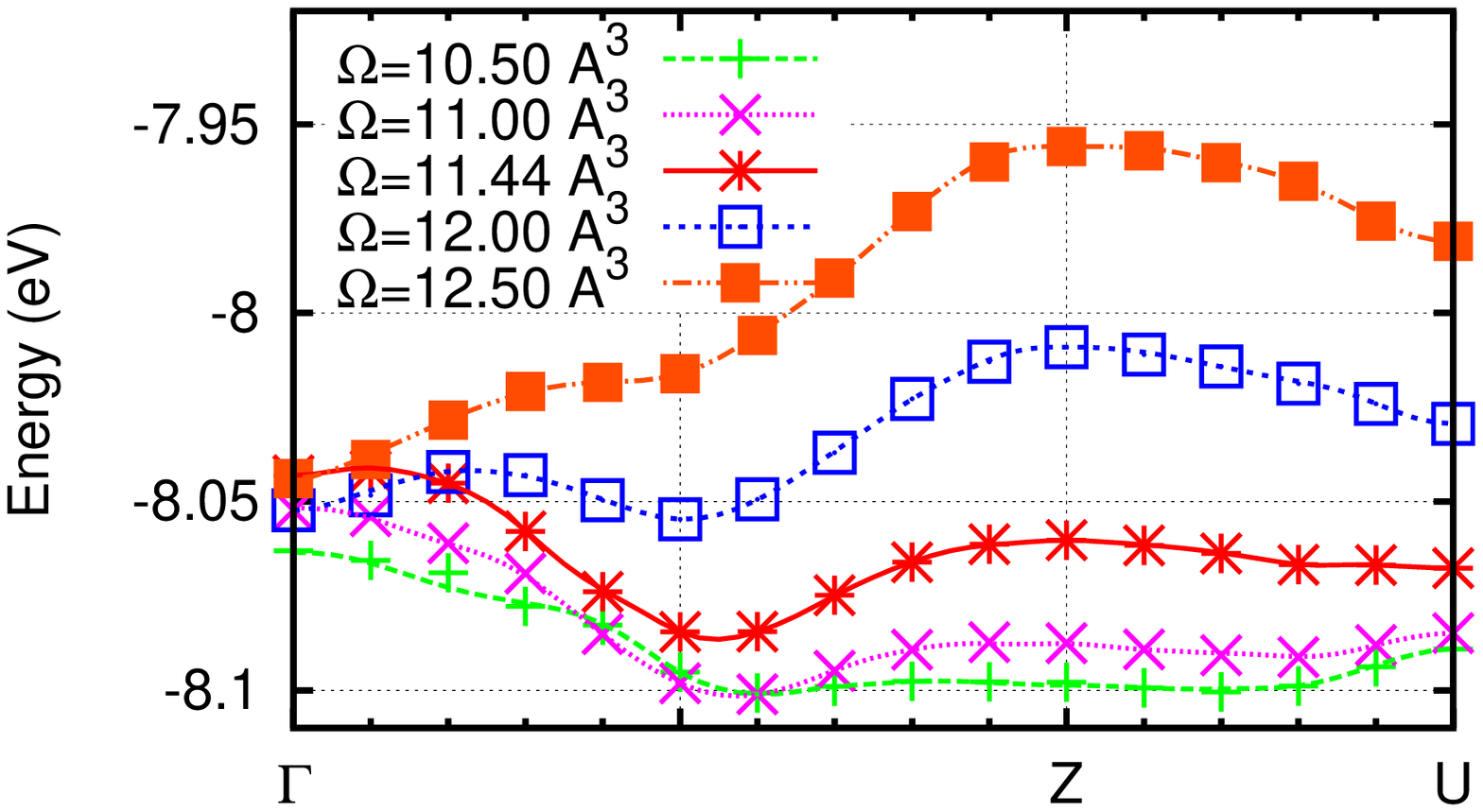}}}\\
\resizebox*{0.98\columnwidth}{!}{\rotatebox{0}{\includegraphics{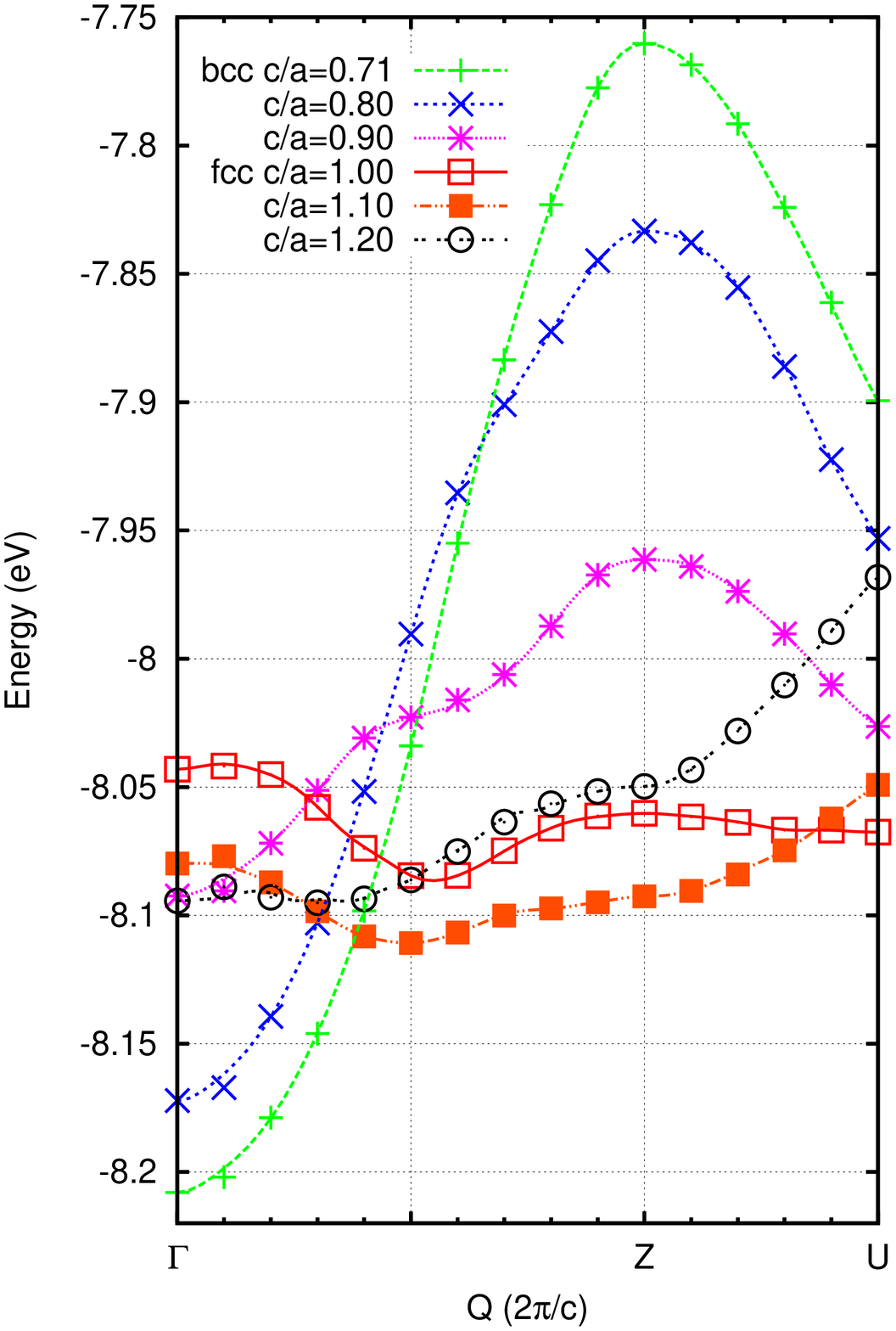}}}
\caption{(Color online) Dependence of the total energy of fcc Fe on the
the spin spiral wave vector $E({\bf Q})$ for different volumes (upper
panel, $c/a$=1) and for different $c/a$ ratios (lower panel,
$\Omega$=11.44~\AA$^3$). The deformations $c/a=1/\sqrt{2}$ and $c/a=1$
correspond to the bcc ($\alpha$-Fe) and fcc ($\gamma$-Fe) structures,
respectively. Symbols mark the results of the density functional
calculations whereas the lines correspond Eq. (\ref{h-ss}) with obtained
exchange parameters $J_n$.}
\label{fig_spiral_e}
\end{figure}

\section{Computational results}

The local magnetic moments $M(\mathbf{Q})$ and total energies for the spin
spiral states $E({\bf Q})$, calculated for different values of volume
and tetragonal deformations are presented in Figs.~\ref{fig_spiral_m}
and \ref{fig_spiral_e}, respectively. We show the results only for the
symmetric directions of the wave vector $\Gamma-\text{Z}-\text{W(U)}$
in Brillouin zone parallel to $\langle 001\rangle$ and $\langle
012\rangle$ in lattice with cubic (tetragonal) symmetry \cite{hafner}.

The magnetic moments depend strongly on the spin spiral wave vector ${\bf Q}$,
as one can see from Fig.~\ref{fig_spiral_m}.
This fact confirms the non-Heisenberg character of magnetic
interactions in iron.
The magnitude of magnetic moments gradually
decrease by about 30\% along $\Gamma-\text{Z}$ direction for all
considered structures except for fcc iron at small volume
$\Omega=11.0$~\AA$^3$. Large difference of magnetic moments for
fcc Fe at $\Gamma$-point at a small volume ($\Omega=11.0$~\AA$^3$) and
bigger ones  ($\Omega \geq 11.44$~\AA$^3$) results from a well
known magnetovolume instability which was discussed in the context
of the Invar problem \cite{Wasserman}.

According to our results (Fig.~\ref{fig_spiral_e}) the ground state of
fcc iron is spin spiral with ${\bf Q}$ varying nearby $0.5\langle
001\rangle$ (in $2\pi/c$ units) with volume and $c/a$ ratio for a broad
interval $10.5<\Omega<12.0\text{~\AA}^3$. The magnetic ground state of
fcc iron is a controversial issue up to now. The antiferromagnetic
double layer structure (AFMD), equivalent the spin spiral with
$0.5\langle 001\rangle$ has been discussed in a series of papers
\cite{antropov-kats,entel99,friak}. The later publications
\cite{hafner,Abrikosov,Tsetseris} show, rather incommensurate ground
states with $\mathbf{Q}$-vector depending on lattice parameters.
Our result are in agreement with the recent calculations
\cite{Abrikosov,hafner,shallcross,mryasov,garcia,Tsetseris}.

An increase of iron volume further $\Omega > 12.0\text{~\AA}^3$ results in the
transition from spin spiral to ferromagnetic (FM) structure
(Fig.~\ref{fig_spiral_e}a). The energy difference $\Delta E_M$ between
FM and antiferromagnetic (AFM) states (or spin spiral structure with ${\bf
Q} = \langle 001 \rangle$) gives a scale of the exchange interaction
energy which decreases monotonously with increasing of the volume  and
finally changes the sign near $\Omega_\text{exp}= 11.44$~\AA$^3$. This volume
corresponds to an experimental value for precipitates of $\gamma$-Fe in
Cu at low temperatures \cite{tsunoda}.

Our results  demonstrate that the magnetic
structure of fcc iron is strongly dependent on the lattice deformations
(Fig.~\ref{fig_spiral_e}).
This conclusion agrees well with the previous investigations of iron
\cite{hafner,Tsetseris,okatov}. In particular, the spin spiral ground
state is changed to the ferromagnetic one within the tetragonal
deformation region along the Bain path from the fcc ($c/a=1$) to bcc iron
($c/a=1/\sqrt{2}$). A magnetic transition to the FM state and its role in
the martensitic transformation have been discussed earlier in
Ref.~\onlinecite{okatov}. In the opposite case, when $c/a >1$, the
tetragonal deformation leads to a weaker dependence of
$E({\bf Q})$. The spin spiral structure represents a ground state
 at $c/a \approx 1.1$ and a transition to the
ferromagnetic ground state appeared at $c/a \geq 1.2$. This magnetic
transformations are in agreement with a previously obtained phase diagram \cite{Tsetseris}.

\begin{figure*}[!thb]
\resizebox*{\textwidth}{!}{\rotatebox{270}{\includegraphics{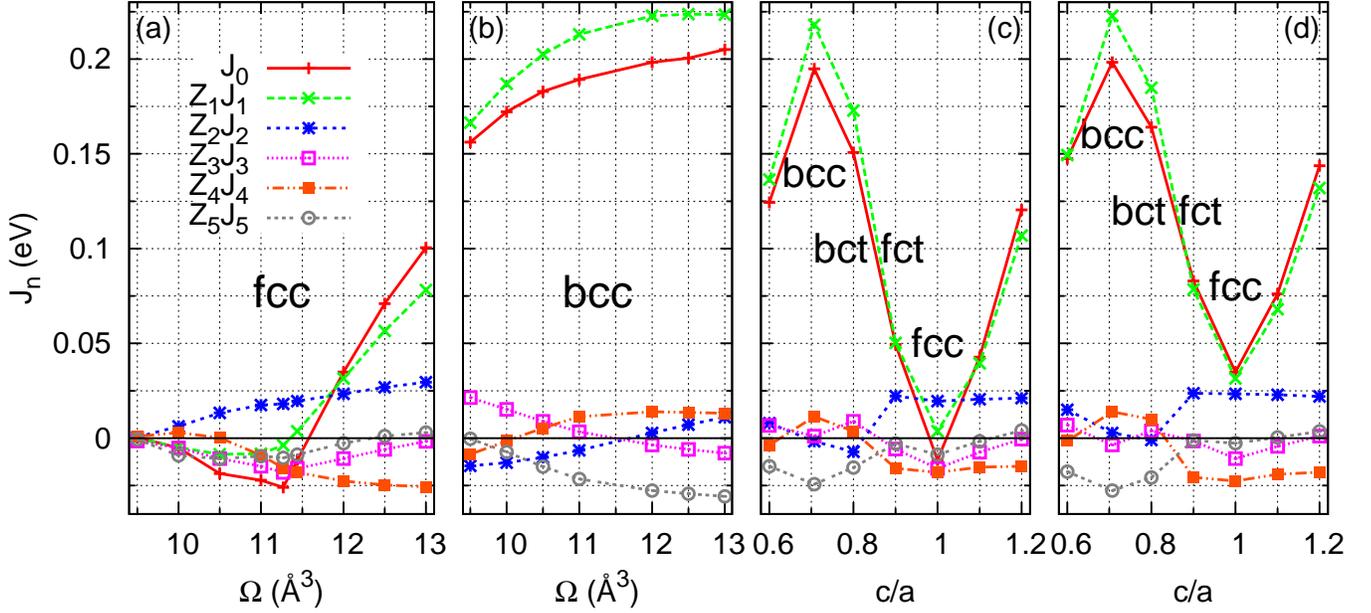}}}
\caption{(Color online) Exchange parameters $J_n$ for $n=1,2,3,4,5$ for
different lattice parameters: dependence $J_n$ on a volume of fcc
(a) and bcc (b) Fe; dependence $J_n$ on $(c/a)$ at fixed volumes
$\Omega=11.44$~\AA$^3$ (c) and $\Omega=12.0\text{~\AA}^3$ (d), respectively.
}
\label{fig_exch}
\end{figure*}

The results for exchange parameters  $J_{n}(c/a,\Omega)$ as function of
lattice distortions are presented in Fig.~\ref{fig_exch}.
Positive values indicate that the ferromagnetic
type of ordering is preferable. The dependence $E({\bf Q})$
determined by Eq. (\ref{h-ss}) with obtained exchange parameters $J_{n}$
give a perfect interpolation to the calculated spin spiral energies
(lines and symbols in Fig. \ref{fig_spiral_e}).
A striking feature of this curves is that the total exchange energy $J_0$ behaves
similar to $Z_1J_1$ ($Z_i$ corresponds to number of $i$-th neighbors )
for all deformations considered. This means that the
contributions of longer-range exchange interactions ($n>1$) are
canceled out. Similar results have been obtained by
analytical calculations of exchange parameters \cite{Ruban} for the volume variation of fcc iron.

Effects of volume variation on the exchange parameters in fcc
structure is very noticeable and $J_1$ demonstrates there a
non-monotonous behavior (Fig.~\ref{fig_exch}a). At low volumes
($\Omega< \Omega_\text{exp}$) total exchange parameter $J_0$ become negative
showing the tendency to antiferromagnetic--type coupling. For
atomic volumes near $\Omega _\text{exp}$ the parameter $J_1$ is
close to zero and the exchange energy $J_0$ is small and negative.
In this case the value $J_0$ is determined by all exchange parameters
$J_n$ with $n>1$. Therefore, computational results for $\Omega
\approx \Omega _\text{exp}$ appear to be quite sensitive to the
details of the approximation used \cite{Abrikosov} (e.g. the exchange-correlation
functional, energy cut-off, number of k-points, etc.).
Such behavior of exchange parameters can likely be related to a complex magnetic
structure discussed in the experimental work by Tsunoda and
co-workers \cite{tsunoda}.

Parameter $J_0$ changes the sign at a volume which is just slightly
above the $\Omega_\text{exp}$ and grows rapidly, therefore the ferromagnetic
order becomes more stable for higher volumes (Fig.~\ref{fig_exch}a).
The behavior of $J_0$ (Fig.~\ref{fig_spiral_e}) agree well with previous calculations
\cite{Ruban,mryasov2} and reproduces the transition from
spin spiral to ferromagnetic state discussed above.

In the bcc Fe exchange parameters demonstrate a rather weak
sensitivity to the volume variation and the nearest neighbor
contribution $J_1$ is large, positive and dominant in a broad
interval of $\Omega$ (Fig.~\ref{fig_exch}b). As a result, the
ferromagnetic ground state has an essential preference in bcc Fe
in comparison with the AFM and non-collinear magnetic structures.
Results of previous calculations
\cite{liecht,mryasov,moran,pajda,schilf,antropov,spisak,frota,rosen}
give slightly lower values $J_0$ and $J_1$ in
bcc Fe than obtained here but also reproduce a dominate contribution
of $J_1$ to the exchange energy.

A dependence of exchange parameters on the tetragonal
deformation $c/a$ is presented in the Fig.~\ref{fig_exch}(c,d).
A symmetry break caused by tetragonal deformations leads to a
modification of the coordination numbers in fcc or bcc lattice.
Here we neglect the rearrangement of site positions and assume
that the set of atoms belongs to the same coordination shells $n$
in fcc and fct structures for $0.85 \leq c/a \leq 1.2$ and in bcc
and bct structures for $0.6 \leq c/a < 0.85$. The curves
$J_1(c/a)$ and $J_0(c/a)$ have both a minima for fcc and a maxima for
the bcc structures. One can see that near the bcc structure $J_0$ is much
less sensitive to the dilatation than to the tetragonal
deformation. Near the fcc, $J_0$ is very sensitive to both types of
deformations. This is mainly due to sensitivity of $J_1$ to deformations whereas
$J_n$ for $n>1$ are almost unchanged  with variation of
lattice parameters.

\begin{figure}
\resizebox*{\columnwidth}{!}{\rotatebox{270}{\includegraphics{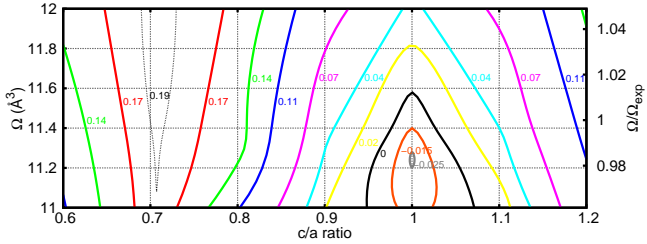}}}
\caption{(Color online) Dependence of the total exchange parameter $J_0$ on
volume $\Omega$ and $c/a$
ratio as a contour plot $J_0(\Omega, c/a)$.}
\label{fig_cont_plot}
\end{figure}

The dependence of the exchange parameters on both types of
deformations is shown in Fig.~\ref{fig_cont_plot} as a contour
plot $J_0(\Omega, c/a)$. One can see that the tetragonal
deformation together with the increase in volume enhance
significantly the exchange interaction energy in $\gamma$-Fe. The
value $J_0 \approx 70$~meV is reached for the experimental volume
of $\gamma$-Fe $\Omega \approx 12$~\AA$^3$ and $(c/a-1) \approx
5\%$.

\begin{figure}
\resizebox*{\columnwidth}{!}{\rotatebox{270}{\includegraphics{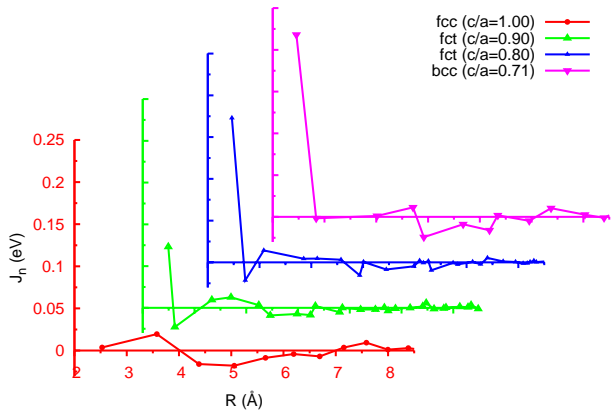}}}
\caption{(Color online) The exchange parameter as a function of
interatomic distance to the $n$-th neighbour $J_n(R_n)$ for different
$c/a$ ratios.}
\label{fig_J(r)}
\end{figure}

Calculated exchange parameters $J_n$ are presented in  the Fig.
\ref{fig_J(r)} as functions of interatomic distances. The
exchange interactions in fcc iron have a very long-ranged behavior
at the volumes $\Omega \approx\Omega_\text{exp}$- Such a strong Friedel
oscillations was already found in Ref.~\onlinecite{Ruban}. This is
a reason of magnetic frustrations and existence of numerous complex
magnetic structures with low energies in the fcc Fe
\cite{Ruban,tsunoda,abrikos,shallcross}. A tetragonal deformation of
the fcc structure changes dramatically the behavior of $J_n$ due to a
sharp increase of $J_1$ contribution which becomes a dominant one. The
increase of volume acts in a similar way. One can see from
Fig.~\ref{fig_J(r)} that the exchange interactions depends not only on interatomic distance $R_n$
but also very sensitive to particular values of $c/a$. Therefore, correct lattice deformations
should be necessarily taken into account {\it explicitly} for a correct
description of magnetic structures in Fe.

\section{Discussion and conclusions}

\begin{figure*}
\resizebox*{\columnwidth}{!}{\rotatebox{270}{\includegraphics{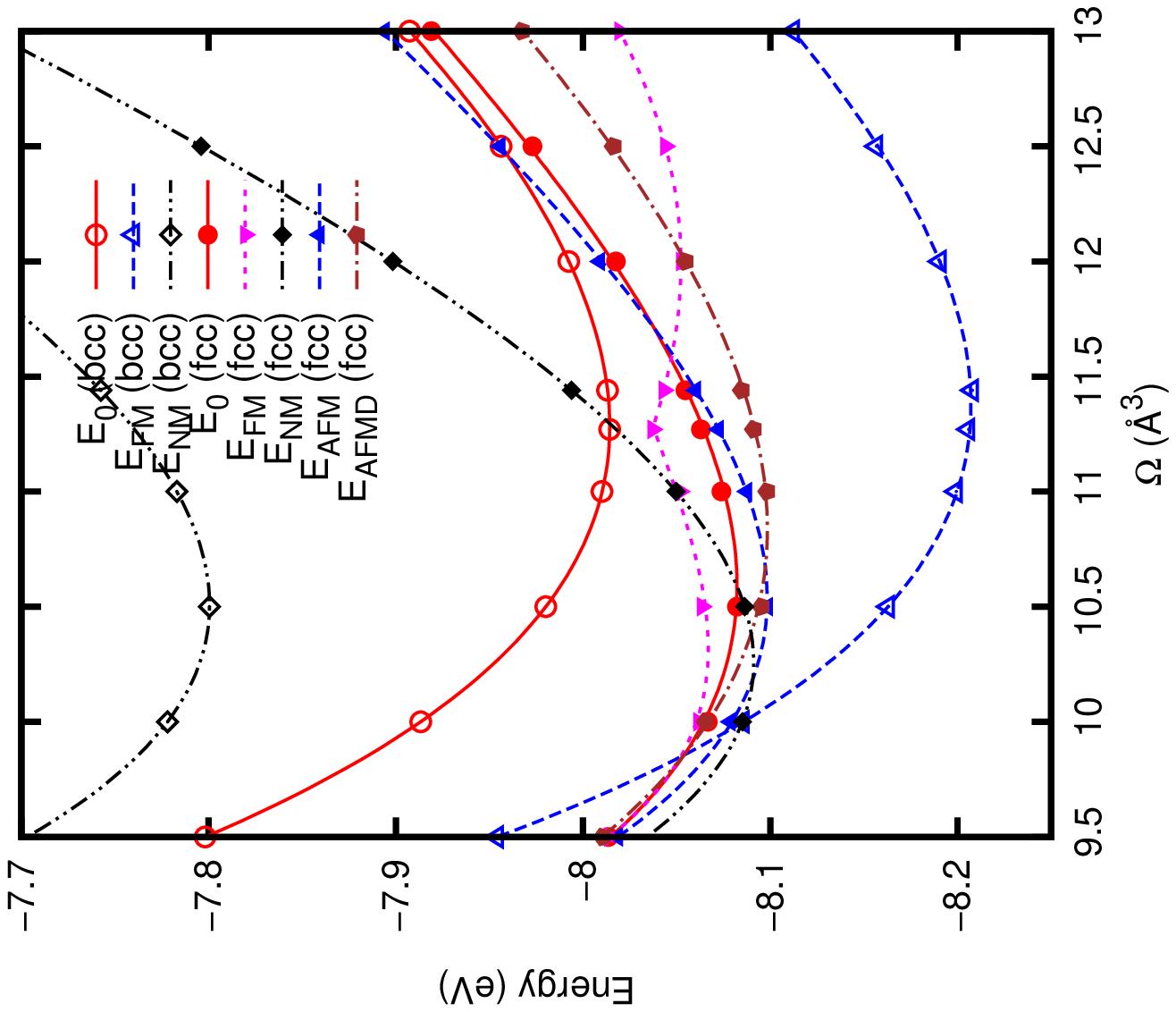}}}
\resizebox*{\columnwidth}{!}{\rotatebox{270}{\includegraphics{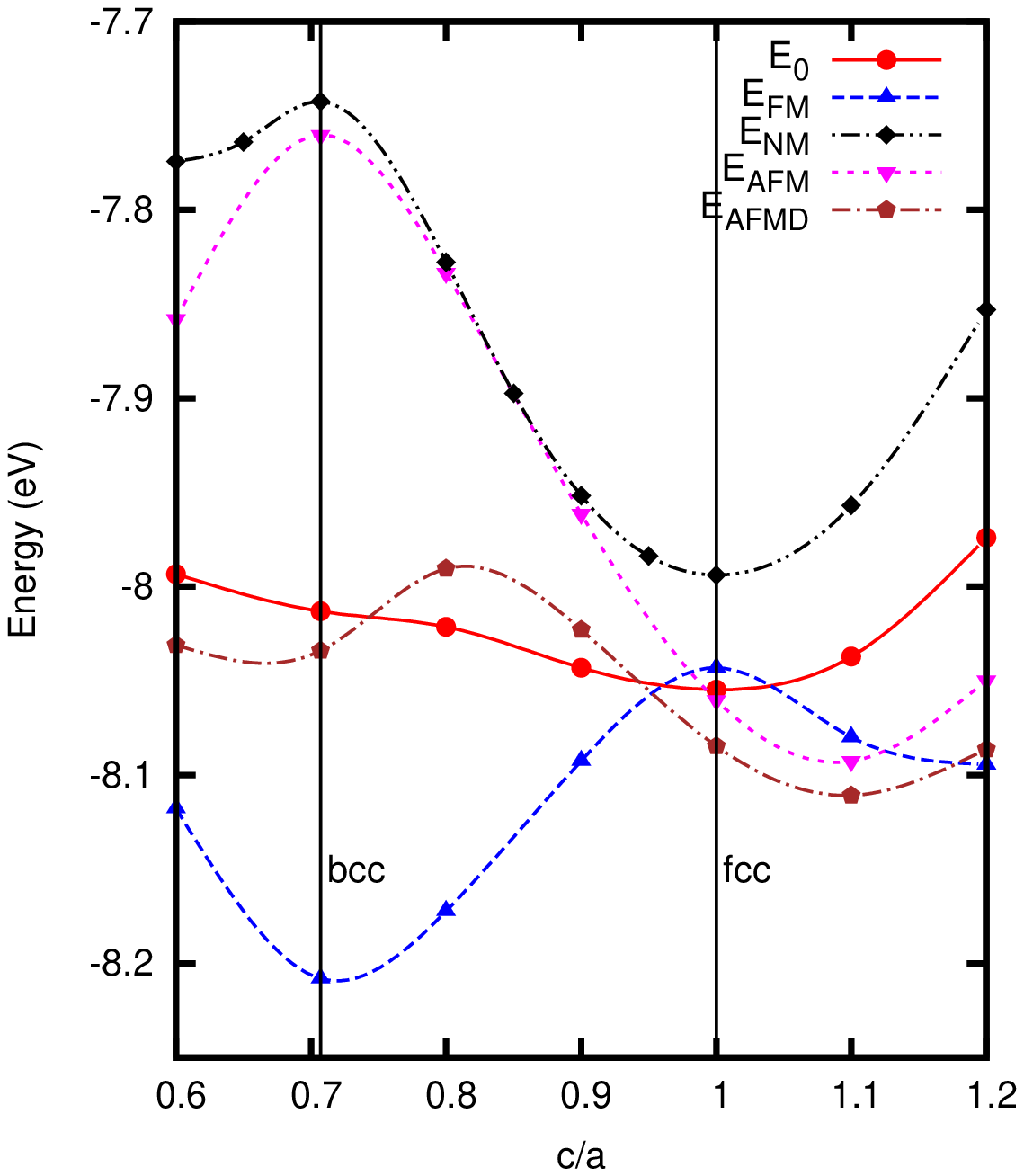}}}
\caption{(Color online) Total energy of iron reconstructed from the $E_0$
and $J_n$ parameters (symbols) as a function of volume (a) and tetragonal
deformation (b) for FM and AFM magnetic structures, as well as the
magnetic-structure-independent contribution $E_0$ to the total energies.
The lines are total energy fitting using Birch-Murnaghan equation of
state\cite{birch} for (a) and splines for (b). Fitting for the FM state in
fcc structure (a) is done separately for each minimum.}
\label{fig_birch}
\end{figure*}

We can determine the magnetic-structure independent contribution
$E_0$ by subtracting the Heisenberg-like contribution with
calculated exchange parameters from the total energy.
In order to do this one can use the energy of ferromagnetic state in the spin-spiral
framework $Q=0$ from the Eq.~(\ref{h-ss}) and the following
expression: $E_0=E_\text{FM}-J_0$. As was mentioned earlier, $E_0$ essentially
differs from a total energy obtained in the non-spin-polarized
calculations because of implicit dependence of $E_0$ on the
magnetic moment $M$. They are equal only for the systems with zero
magnetic moments of all atoms.

The results for $E_0$ are shown in Fig.~\ref{fig_birch} together with
the total energies of FM bcc and FM, AFM and AFMD fcc states obtained by
the reconstruction from $E_0$ and $J_n$. For comparison, the total
energies of $E_\text{NM}$ obtained from calculations by VASP are also
shown. These results agree very well with the previous {\it ab-initio}
calculations\cite{friak} and demonstrate the dramatic difference between
$E_0$ (see Eq.~\ref{E0}) and $E_\text{NM}$.

For fcc iron the magnetic-structure independent
contribution $E_0$ is rather close to the energy of
AFM and AFMD states. For bcc iron the difference between $E_0$ and
ground-state energy $E_\text{FM}$ is larger but rather weakly
volume dependent compare to fcc states. At the same time, the
energy of FM fcc state shows two minima at low and high volumes.
This behavior of fcc total energy drastically differs from the $E_0$ curve. The
difference is larger for higher volumes and has entirely magnetic
origin due to increase of the exchange parameters with $\Omega$
(Fig.~\ref{fig_exch}). Quantitatively, the values of bulk modulus
for fcc iron obtained from the Birch-Murnaghan equation of
state\cite{birch} for $E_0$ and $E_\text{AFMD}$, $E_\text{AFM}$
curves differ by about 17\% and 30\% (161, 189, and 207 GPa,
respectively). For the bcc iron estimation of bulk modulus from $E_0$ and
$E_\text{FM}$ curves give the same $B \approx 187$ GPa, in
agreement with the experiment \cite{BbccFe}.

The situation with tetragonal deformations is quite unusual.
One can see in Fig.~\ref{fig_birch}b that $E_0$ depends on
$c/a$ very weakly. This means that the Heisenberg-like contribution is
dominant in the shear modulus $C^{\prime}$, as well as in the whole energy
curve along the Bain path. This is main origin of anomalously strong
coupling between the magnetic and lattice degrees of freedom in iron,
where the tetragonal deformation plays a special role. The curve
$E_0(c/a)$ has a minimum at $c/a=1$ (fcc structure) whereas both
$E_\text{FM}$ and $E_\text{AFM}$ have no minima at this point which
means instability of fcc phase in both magnetic structures. The minima
correspond to bcc (FM) and fct (AFM, AFMD) states with $c/a>1$.

Our calculations reveal another unusual feature of the magnetic interactions in fcc iron
related with a growth of the exchange parameter $J_1$ and, as
a consequence, $J_0$ with the volume increase at $\Omega >
\Omega_\text{exp}$ (see Fig.~\ref{fig_exch}a). This behavior corresponds
to the rising branch of the Bethe-Slater curve $J (\Omega)$ which have
been used for a semi-quantitative interpretation of the Invar anomaly
\cite{bethe-slatter}. This region of volumes corresponds to observed
high-temperature phase of $\gamma$-Fe; for further increase of
interatomic distances the overlap of $d$-orbitals becomes weaker and the
exchange interactions $J_n$ decreases.

Here we show that the calculated dependence of $J_n (\Omega)$ can explain
the anti-Invar phenomenon in $\gamma$-Fe \cite{Acet}. If a magnetic
subsystem is well described by the Heisenberg-like model
(\ref{heisen}) its contribution to pressure according to the
Hellman-Feynman theorem is
\begin{eqnarray}
P_m && = -\frac{1}{N}\left\langle\frac{\partial H_{ex}}{\partial
\Omega}\right\rangle= \frac{1}{N}\sum_{i<j} \frac{\partial J_{i,j}}{\partial
\Omega}\langle {\bf e}_i{\bf e}_j \rangle
\nonumber \\
&& \approx Z_1 \frac{\partial J_{1}}{\partial \Omega}\langle {\bf
e}_0{\bf e}_1 \rangle
\label{Pm}
\end{eqnarray}
where $Z_1$ is the number of nearest neighbors. We assume that the
nearest-neighbor interaction is the strongest one which is supported
by our fist-principle calculations.
For a purpose of qualitative discussions, we will treat exchange interactions perturbatively assuming
$Z_1 J_1 \ll T$ ($T$ is the temperature and $k_B=1$).
Then one has $\langle {\bf e}_0{\bf e}_1
\rangle \approx J_1/3T$ and therefore
\begin{equation}
P_m=\frac{Z_1}{6T}\frac{\partial J_{1}^2}{\partial \Omega}
\label{Pm_approx}
\end{equation}
This means that the pressure induced by magnetic exchange interactions
is positive and decreases with the temperature increase.

The thermal expansion coefficient
\begin{equation}
\alpha = \frac{1}{\Omega} \left(\frac{\partial \Omega}{\partial
T}\right)_P = \left (\frac{1}{B} \right)_T \left(\frac{\partial P}{\partial T}\right)_{\Omega}
\label{alpha_t}
\end{equation}
can be divided into magnetic-structure independent part ($\alpha_0$) and
one related with magnetic exchange interactions ($\alpha_m$): $\alpha =
\alpha_0+\alpha_m$. The magnetic exchange part is equal to
\begin{equation}
\alpha _m = \frac 1 B \left(\frac{\partial P_m}{\partial T} -B_m \alpha _0 \right)
\label{alpha_m}
\end{equation}
Here $B$ is isothermal bulk modulus $B=B_0+B_m$, $B_0$ is the
magnetic-structure independent part of $B$, and $B_m = -\Omega
({\partial P_m}/{\partial \Omega})_T$. Usually, the second term in
Eq.(\ref{alpha_m}) is neglected. Since ${\partial P_m}/{\partial
T} <0$ one can assume that the expression (\ref{Pm_approx}) should
lead to the Invar behavior, and hence to the negative contribution
to the thermal expansion coefficient. A strong volume
dependence of the exchange parameter $J_1$ can leads to the
opposite conclusion. Substitutes Eq. (\ref{Pm_approx}) into Eq.
(\ref{alpha_m}) one finds
\begin{equation}
\alpha_m= -\frac{1}{B_0}\frac{Z_1}{6T}\frac{\partial J_{1}^2}{\partial \Omega}
\left[1-\alpha_0 T \frac{\partial \ln (\partial J_1^2/\partial \Omega)}
{\partial \ln \Omega}\right]
\label{alpha_m2}
\end{equation}
Using calculated volume dependence $J_1(\Omega)$ (Fig. \ref{fig_exch})
and values $\alpha_0$, $\Omega$ obtained from the experiment
\cite{Acet} one can find that the second term in square brackets
in the right-hand side of Eq. (\ref{alpha_m2}) is approximately 1.1 at
the temperature of 1200 K. Therefore, a total magnetic exchange contribution to the thermal
expansion coefficient (\ref{alpha_m2}) has {\it positive} sign. This
corresponds to the {\it anti}-Invar behavior, in a qualitative agreement
with the experimental data \cite{Acet}.

The negative magnetic exchange
contribution to the thermal expansion coefficient $\alpha _m$ in the Invar
materials usually is associated with a thermal dependence of the spontaneous
magnetostriction \cite{Mohn-Khmelevski}, while the positive
contribution (anti-Invar behavior) is often considered to be related to
thermal volume changes due to magnetic fluctuations
\cite{Acet,Entel}. The present investigation allows us to explain
the anti-Invar effect of high-temperature $\gamma$ phase of iron within
a simple Heisenberg-like model in terms of magnetic softening of the bulk modulus,
without any assumptions about two magnetic states of iron atoms
with high and low volumes \cite{Wasserman}.

Due to the thermal expansion effective exchange parameters increases
with the temperature increase,
\begin{equation}
J_0^{ef}=J_0+\lambda T,
\label{Jef}
\end{equation}
with a positive constant $\lambda >0$. If we substitute this formula into the
mean-field expression for the magnetic susceptibility
\cite{Vonsovsky1974,exchange,liecht},
\begin{equation}
\chi = \frac{m^2}{3(T-2J_0^{ef}/3)}
\label{susc}
\end{equation}
one can see that corresponding temperature dependence leads to an increase of the effective magnetic
moment, $m ^2 \rightarrow m^2/(1-2\lambda /3)$ and the Curie temperature,
$T_C \rightarrow T_C/(1-2\lambda /3)$.

To conclude, we have carried out a systematic study of exchange
parameters in $\alpha$- and $\gamma$-Fe as functions of the volume
and tetragonal deformation. The computational results
demonstrate a strong coupling between lattice and magnetic degrees
of freedom which should be taken into account in
thermodynamic properties of Fe, especially its thermal expansion.
Accurate analysis of the magnetic-structure independent contribution
$E_0$ allows us to conclude that a response of fcc and bcc Fe to deformations
is mainly controlled by the magnetic exchange.

\section{Acknowledgments}

M.I.K. acknowledges financial support from EU-Indian scientific collaboration
program, project MONAMI. We thank Igor Abrikosov for critical fruitful discussions.
The calculations were partly performed on the supercomputer at NRC ``Kurchatov
Institute''.

{}

\end{document}